%% file: main.tex
\documentclass[a4paper, amsfonts, amssymb, amsmath, reprint, showkeys, nofootinbib, twoside, superscriptaddress]{revtex4-1}
\usepackage[english]{babel}
\usepackage[utf8]{inputenc}
\usepackage{comment}
\usepackage{siunitx}
\usepackage[hidelinks]{hyperref}
\usepackage[colorinlistoftodos, color=green!40, prependcaption]{todonotes}
\input{preamble}
\begin{document}
\title{Free-space multipass optical parametric amplifier}

\author{Milan Fayat}
    \email[Correspondence email address: ]{milan.fayat@institutoptique.fr}
    \affiliation{Université Paris-Saclay, Institut d’Optique Graduate School, CNRS, Laboratoire Charles Fabry, 91127 Palaiseau, France}
\author{Giovanni Cichelli}
    \affiliation{Université Paris-Saclay, Institut d’Optique Graduate School, CNRS, Laboratoire Charles Fabry, 91127 Palaiseau, France}
\author{Michele Natile}
    \affiliation{Amplitude, 11 Avenue de Canteranne, Cité de la Photonique, 33600 Pessac, France}
\author{Yoann Zaouter}
    \affiliation{Amplitude, 11 Avenue de Canteranne, Cité de la Photonique, 33600 Pessac, France}
\author{Marie Ouillé}
    \affiliation{Université Paris-Saclay, Institut d’Optique Graduate School, CNRS, Laboratoire Charles Fabry, 91127 Palaiseau, France}
\author{Patrick Georges}
    \affiliation{Université Paris-Saclay, Institut d’Optique Graduate School, CNRS, Laboratoire Charles Fabry, 91127 Palaiseau, France}
\author{Marc Hanna}
    \affiliation{Université Paris-Saclay, Institut d’Optique Graduate School, CNRS, Laboratoire Charles Fabry, 91127 Palaiseau, France}
\author{Xavier Délen}
    \affiliation{Université Paris-Saclay, Institut d’Optique Graduate School, CNRS, Laboratoire Charles Fabry, 91127 Palaiseau, France}
\date{\today} 

\begin{abstract}
Scaling the efficiency of optical parametric amplifiers (OPAs) without degrading spatio-temporal pulse quality is fundamentally limited by spatio-temporal walk-off, intensity dependent gain, and back-conversion. Here, we numerically and experimentally demonstrate an OPA architecture based on a free-propagating quasi-periodic geometry that overcomes these bottlenecks. Operating within a single nonlinear crystal, the system utilizes pass-by-pass dichroic idler rejection to suppress back-conversion, a birefringent crystal for temporal resynchronization, and free-space diffraction to improve the spatial overlap along propagation. Starting from 1.9 µJ 330 fs pulses at 515 nm and a continuous-wave seed at 783 nm, the generation of 0.8 µJ 160 fs  signal pulses at the same wavelength  is obtained at a repetition rate of 500 kHz. This simple architecture achieves a 64\% quantum efficiency, a 42\% pump-to-signal conversion efficiency, and 80 dB of gain while maintaining excellent spatial and temporal quality, providing a scalable platform for ultrafast sources with arbitrary emission wavelengths.
\end{abstract}


\maketitle

Femtosecond optical parametric amplifiers (OPAs) are widely established as indispensable tools across a broad range of scientific disciplines. Pumped by femtosecond lasers, they allow to generate ultrashort pulses widely tunable at wavelengths ranging from the ultraviolet to the mid-infrared \cite{cerullo_ultrafast_2003}. Their use has enabled breakthroughs in time-resolved spectroscopy \cite{maiuri_ultrafast_2020, Luo:16}, strong-field physics \cite{krausz_attosecond_2009,Popmintchev:12}, coherent control of chemical reactions \cite{Hildner:25}, and multiphoton microscopy \cite{wu_dual-wavelength_2022,wang_three-photon_2020,helmchen_deep_2005}. Yet despite decades of refinement, the energy conversion efficiency of femtosecond single-stage OPAs remains fundamentally constrained, typically ranging from 10 to 30\% when accounting for the combined signal and idler outputs for systems aiming at preserving a high beam and pulse quality \cite{manzoni_design_2016}. 

Several physical mechanisms are responsible for this limitation. Temporal and spatial walk-offs between the interacting beams reduce the effective nonlinear overlap along the crystal, degrading both gain and beam quality \cite{cerullo_ultrafast_2003,manzoni_design_2016}. Critically, back-conversion, the reversal of energy flow from signal and idler back to the pump, imposes a limit on conversion efficiency. Together with the gain filtering associated to the pump intensity profile, this process often caps the achievable efficiency and is responsible for the degradation of the spatial and temporal profiles of the amplified wavepacket \cite{Arisholm:04}.

The most widespread strategy to circumvent back-conversion and walk-off relies on adding amplification stages, where each successive stage is pumped by a fraction of the available pump energy and independently optimized. While this sequential approach yields improved conversion efficiency, it comes at the cost of substantial optical complexity resulting in cumbersome, large-footprint setups prone to misalignment. Another approach consists in using nonlinear crystals that absorb the idler beam \cite{Ma:15,Ma2022}; however, its scalability remains constrained by absorption-induced thermal effects.

Inspired by the remarkable success of multi-pass cells (MPCs) for nonlinear pulse post-compression \cite{viotti:22, hanna:2021}, several groups have recently extended this concept to optical parametric amplification \cite{nagele_dispersion-engineered_2025, kadriu_continuous-wave_2026, rajhans_optical_2026}. Despite differences in footprint and pump energy, these systems share a core principle: embedding the nonlinear crystal within an MPC engineered to simultaneously suppress the idler and compensate for the pump-signal group delay after each pass. By effectively inhibiting back-conversion, this approach overcomes the gain-bandwidth trade-off of conventional single-pass configurations while preserving near-diffraction-limited beam quality.

Using stable Herriott or White cell geometries that mimic a free-space waveguide, these multi-pass amplifiers have demonstrated remarkable performance. Pulsed configurations seeded by optical parametric oscillators or supercontinuum sources have achieved pump-to-signal quantum conversion efficiencies of 52\% \cite{nagele_dispersion-engineered_2025} and 65\% \cite{rajhans_optical_2026}, while continuous-wave (CW) seeding has yielded efficiencies around 30\% \cite{kadriu_continuous-wave_2026}. Implementation strategies have converged into two main designs: compact, folded cells utilizing specialized dispersion-compensating mirrors for passive spatial and temporal overlap \cite{nagele_dispersion-engineered_2025, kadriu_continuous-wave_2026}, and independent multi-mirror arrangements that minimize insertion losses at the expense of a complex pass-by-pass alignment procedure \cite{rajhans_optical_2026}.

Here, we introduce a new multipass OPA concept solely based on the free space propagation of both pump and signal beams, without any refocusing optics of any kind. Pump, signal, and idler propagate through a succession of gain stages, in a single nonlinear crystal. After each stage, the idler is discarded to suppress back-conversion, and the group delay between pump and signal accumulated in the nonlinear crystal is compensated using a simple birefringent crystal, alleviating the need for custom dispersion-compensating mirrors. Between successive gain stages, rather than imposing a fixed periodic propagation through relay optics, we exploit the natural differential diffraction between pump and signal: since each high gain stage reduces the signal beam size, and because the signal wavelength is longer, they diverge at different rates in free space. This can be used to passively optimize mode matching in successive stages if the free propagating distance between stages is chosen adequately. This differential diffraction fulfills the role of the spatial mode management that MPC approaches achieve through quasi-waveguiding.

Based on (3+1)D numerical simulations to guide the design and understand the underlying physics, we demonstrate an experimental implementation of this simple concept that achieves results that are on par with MPC-based OPAs, with a 64\% pump-to-signal quantum efficiency, excellent temporal and spatial qualities, while achieving an overall gain of 80 dB, which enables CW laser diode seeding. As opposed to seeding methods based on white-light generation, this allows to scale the repetition rate arbitrarily beyond 1 MHz and alleviates the need for precise synchronization between pump and seed \cite{kadriu_continuous-wave_2026,wang_cwseeded_2024}.
\begin{figure}[h]
\centering
\includegraphics[width=0.95\linewidth]{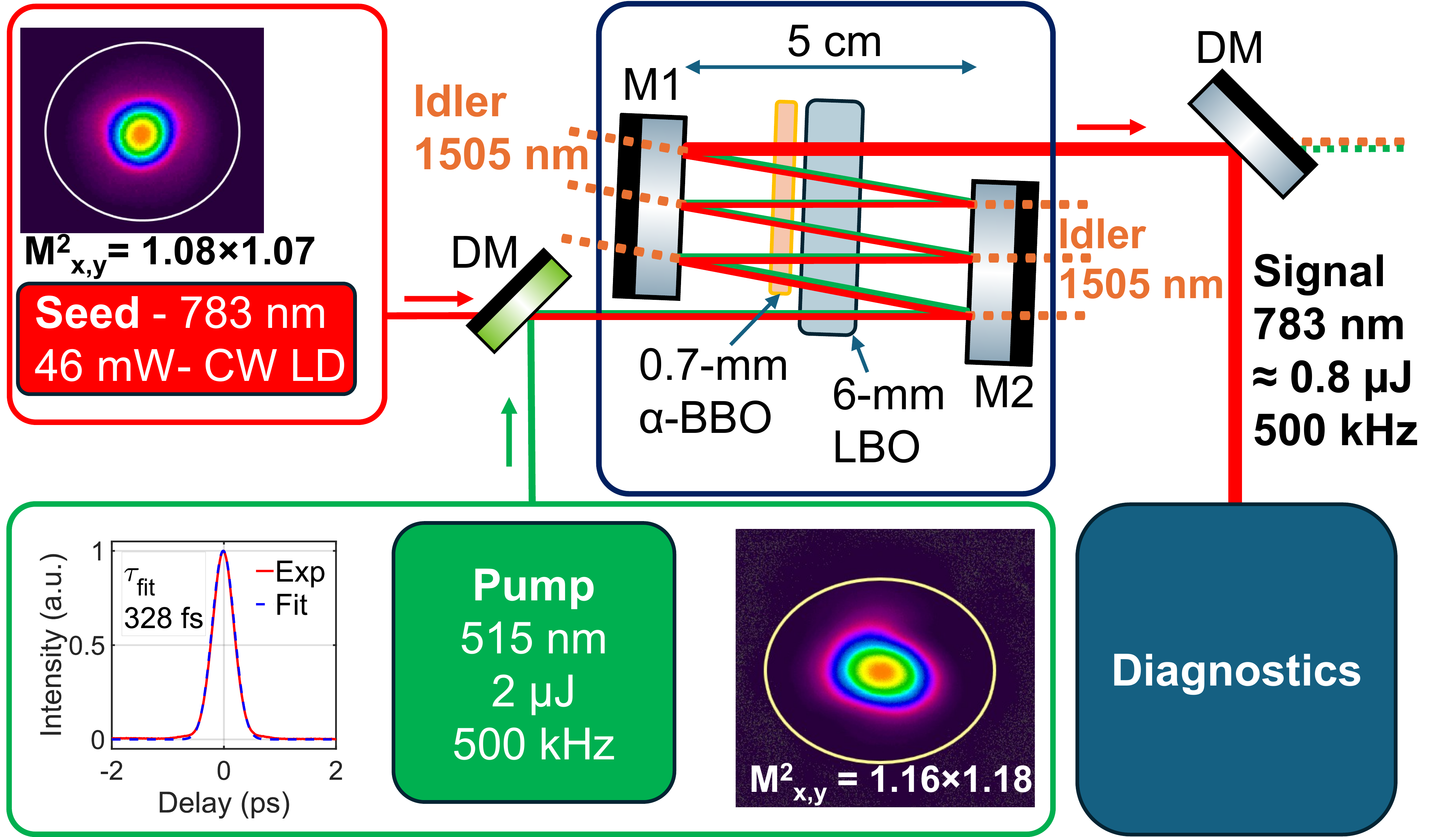}
    \caption{Experimental setup. CW LD: continuous-wave laser diode; M1/M2: HR mirrors at 515 and 783 nm, HT at 1505 nm; DM, dichroic mirror.}
    \label{fig:exp_setup}
\end{figure}

The configuration investigated both numerically and experimentally is schematically depicted in Fig.~\ref{fig:exp_setup}. The pump source is a commercially available frequency-doubled Yb-doped ultrafast fiber laser delivering 2 \si{\micro\joule} pulses of 330 fs duration at 515 nm and a repetition rate of 500 kHz. The OPA signal seed is provided by a single-frequency, single-mode fiber-coupled CW diode laser operating at 783 nm with up to 50 mW of output power. The OPA medium is a 6 mm-long type-I LBO crystal, anti-reflection coated for both pump and signal wavelengths. The crystal length is chosen such that the group-velocity mismatch (GVM) between pump and signal induces a complete temporal walk-through of the amplified signal across the pump pulse envelope over two successive passes through the crystal. In our case, this group delay difference is 400 fs over 12 mm of LBO. Two plane folding mirrors, highly reflective for both pump and signal and highly transmissive for the idler, redirect the beams to realize seven passes through the LBO crystal. These passes are realized in the horizontal plane, while the critical phase matching angle is in the vertical plane. The idler is ejected at each reflection to limit back-conversion. The group delay between the orthogonally polarized pump and signal pulses accumulated after two passes in the OPA crystal is compensated using an AR-coated 700 \si{\micro\meter}-thick alpha-BBO crystal. Pump and signal beams propagate collinearly and are focused to a waist radius of $w_{0, x}^{\mathrm{pump}} = 150$~\si{\micro\meter}, $w_{0,y}^{\mathrm{pump}} = 125$~\si{\micro\meter} and $w_0^{\mathrm{seed}} = 130$~\si{\micro\meter} respectively at the first pass inside the LBO crystal. 

\begin{figure}[h]
\centering\includegraphics[width=\linewidth]{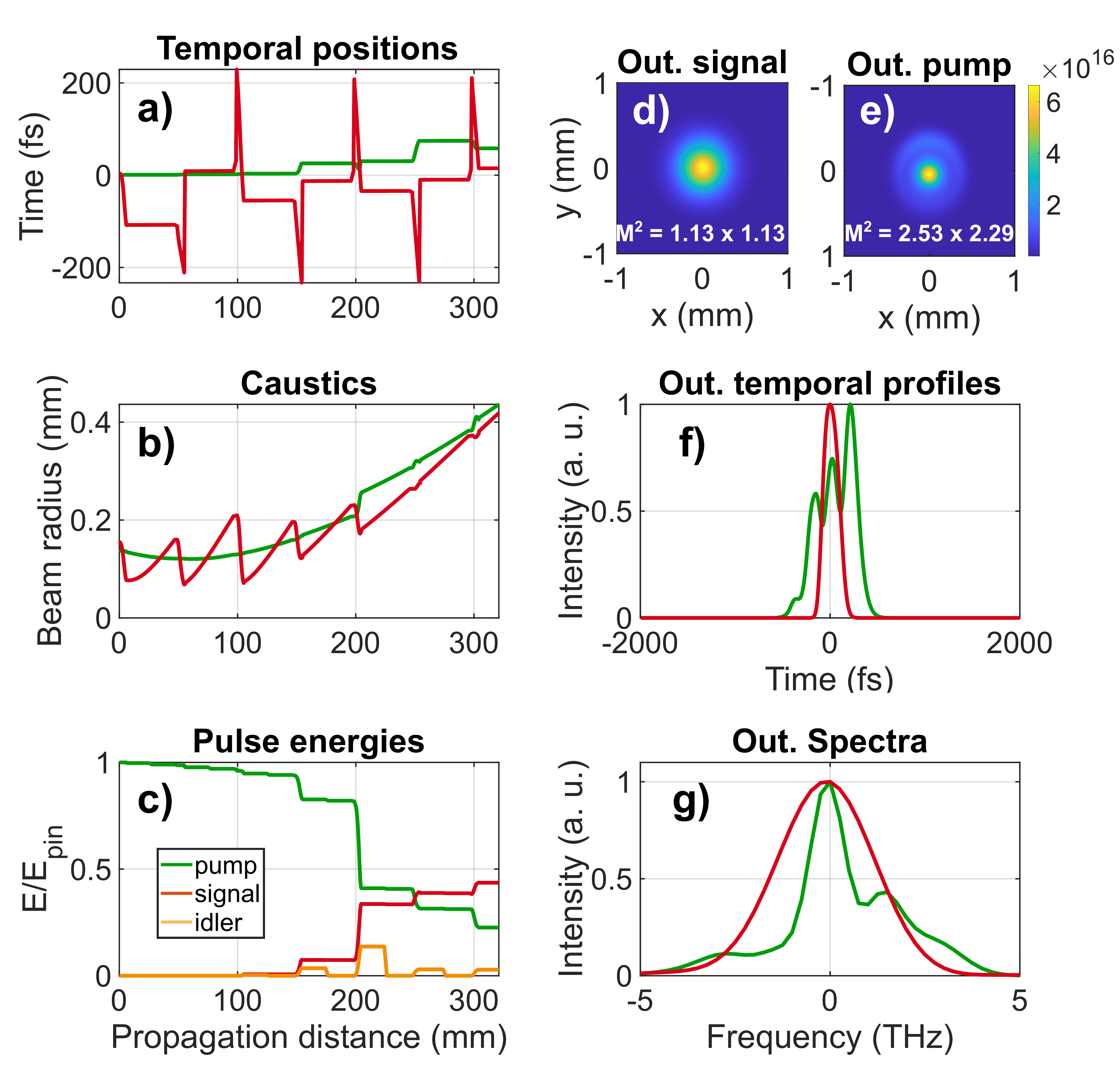}
\caption{Simulated pulse propagation in the multipass OPA for a pump energy of 1.65 \si{\micro\joule}. \textbf{(a)} Temporal positions vs. propagation distance. \textbf{(b)} Evolution of the beam sizes vs. propagation distance. \textbf{(c)} Normalized pulse energies vs. propagation distance. \textbf{(d)} Signal and \textbf{(e)} pump spatial profiles at the output. \textbf{(f)} Normalized temporal profiles at the output. \textbf{(g)} Normalized output spectra.}

\label{fig:simulations}
\end{figure}

To gain quantitative insight into the dynamics of the amplification process and guide the experimental implementation, we perform numerical simulations of the full propagation through the system. They are based on a split-step Fourier algorithm solving the coupled envelope propagation equations for the three interacting fields, in (3+1) dimensions. All relevant linear and nonlinear effects are accounted for: material dispersion is described via the Sellmeier equations appropriate to each crystal and polarization state, wavelength-dependent diffraction is included to faithfully capture the differential beam divergence that underlies the passive mode reshaping between stages, and both second- and third-order nonlinear interactions are incorporated. This model accurately accounts for GVM, spatial and temporal walk-off, and the spatio-temporally varying parametric gain arising from the intensity-dependent nature of the nonlinear interaction. The input spatial beam profile along both axes and temporal pulse shape are assumed to be perfectly Gaussian. In addition, the losses measured experimentally for each optical component are included in the model, ensuring a realistic comparison with the experimental results. The results obtained for the experimental configuration described above are presented in Fig.~\ref{fig:simulations} for a pump energy of 1.65 \si{\micro\joule}. A movie illustrating the propagation with the simulation results is included in the supplementary materials.
Numerical optimization of the free-space propagation distance between successive gain stages reveals that a separation of ~5 cm yields the best beam size adaptation between pump and signal prior to each gain stage, as illustrated in Fig.~\ref{fig:simulations}(b). The physical picture underlying this optimization is the following. In the early passes, the Gaussian intensity profile of the pump imparts a strongly non-uniform gain across the signal beam, with the on-axis region experiencing significantly higher amplification than the wings. This spatial gain narrowing causes an important reduction of the signal beam size upon amplification, progressively degrading the spatial mode-matching between the pump and the signal beams. The inter-stage differential diffraction between pump and signal naturally re-expands and reshapes the signal beam relative to the pump, restoring a favorable spatial overlap at the entrance of the next crystal pass. Furthermore, in the later passes, as the system approaches saturation, pump depletion occurs preferentially at the beam center where the intensity is highest, creating a spatial hole in the pump profile. This is apparent at a propagation distance of $\sim200$ mm (Fig. \ref{fig:simulations}(b)), where the gain stage that converts the largest energy from the pump to the signal causes an abrupt increase in the pump beam size because of this depletion. Left unaddressed, this would severely degrade both the conversion efficiency and the beam quality of the amplified signal. Once again, together with idler suppression, the free-space propagation length plays a critical role, allowing diffraction to smooth out the depleted pump profile between stages and prevent the spatial modulation from being imprinted onto the signal in the last extraction passes. The temporal dynamics of the system is illustrated in Fig.~\ref{fig:simulations}(a), which shows the effectiveness of the periodic GVM compensation imposed by the $\alpha$-BBO plates. Beyond simply compensating the differential group delays, the repeated 400 fs temporal sweeps of the signal across the pump ensure that the signal successively extracts the energy stored across the entire pump pulse duration (330 fs), a critical condition for achieving high extraction efficiency. The combination of the strongly intensity-dependent parametric gain with the quasi-periodic temporal dynamics also has a direct consequence on the output pulse duration: despite being seeded by a CW source, the amplified signal emerges as a pulse of 165 fs duration FWHM, approximately half that of the pump, with a spectral bandwidth of 6.1 nm FWHM corresponding to a time-bandwidth product of 0.495, close to the Fourier limit (Fig.~\ref{fig:simulations}(f,g)). This pulse shortening arises naturally from the temporal gating imposed by the nonlinear gain profile. From a spectral standpoint, the multipass architecture with periodic resynchronization provides a particularly favorable combination. The spectral acceptance bandwidth is governed by the phase-matching conditions of a single crystal pass, preserving the broad acceptance of a short interaction length, while the overall gain is determined by the cumulative nonlinear material traversed across all passes \cite{nagele_dispersion-engineered_2025}. The progressive energy transfer from pump to signal and idler is shown in Fig.~\ref{fig:simulations}(c). After seven passes through the gain crystal, the pump-to-signal power conversion efficiency reaches 44\%, with a corresponding quantum efficiency of 67\%, figures that compare favorably with the best results reported for MPC-based OPA architectures \cite{nagele_dispersion-engineered_2025, rajhans_optical_2026, kadriu_continuous-wave_2026}. Finally, the spatial beam quality of the amplified signal, evaluated from the simulations, yields M$^{2}$ values of $1.13 \times 1.13$, confirming that the passive spatial reshaping mechanism described above successfully prevents the non-uniform gain and pump depletion from degrading the output beam profile, shown in Fig.~\ref{fig:simulations}(d). The output spatial, temporal, and spectral profiles of the pump are all heavily distorted because of the very large depletion (Fig.~\ref{fig:simulations}(e,f,g)).

\begin{figure}[h]
\centering\includegraphics[width=0.95\linewidth]{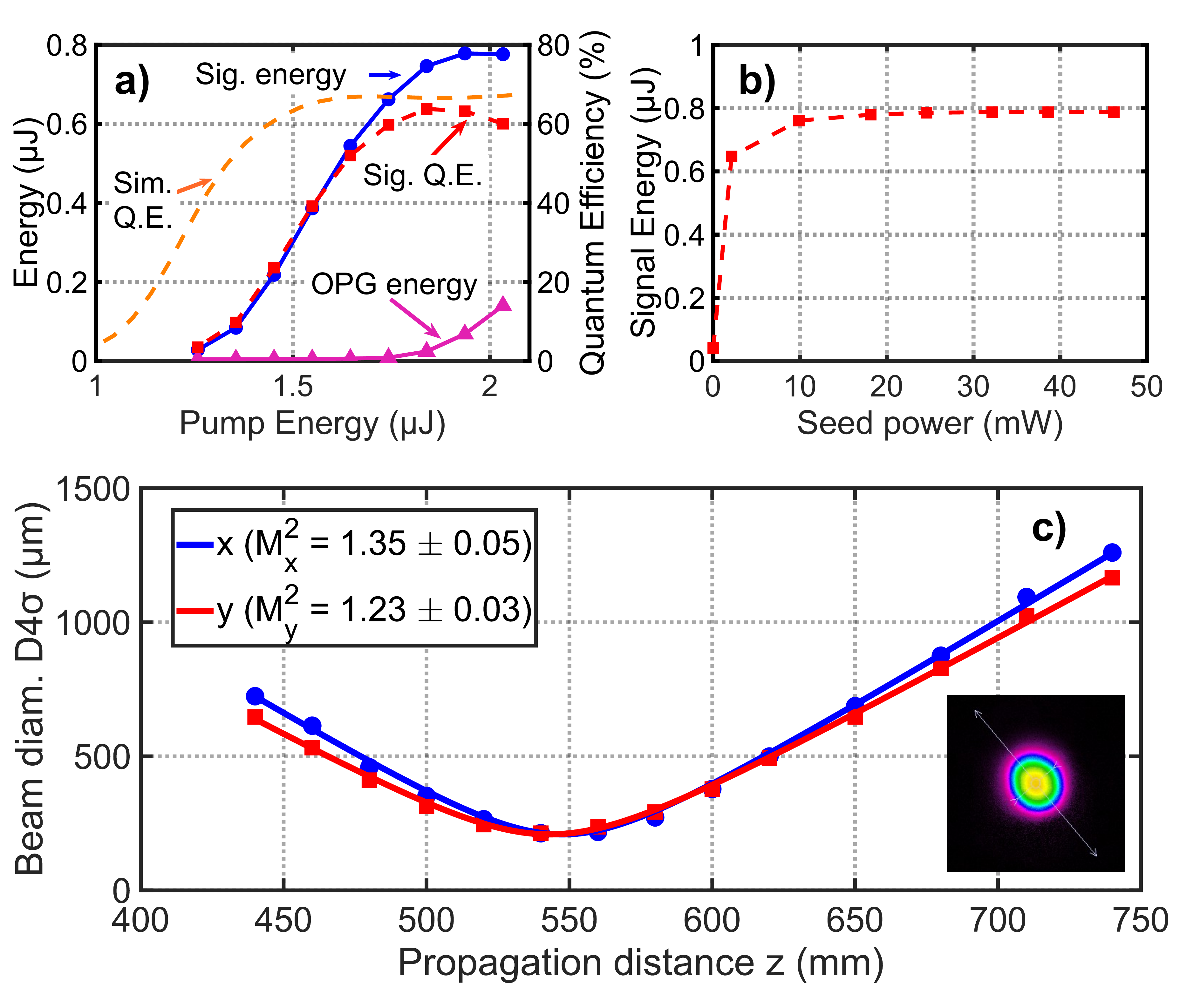}
\caption{\textbf{(a)} Signal energy (blue circles, left axis) and OPG energy (purple triangles, left axis) as a function of pump energy. OPA quantum efficiency (red squares, dashed line, right axis) and simulated quantum efficiency (orange dashed line, right axis) versus pump energy.
\textbf{(b)} Signal energy (red squares, dashed line) as a function of seed power.
\textbf{(c)} Caustic measurements of the 783 nm signal beam along the $x$ (blue circles) and $y$ (red squares) transverse directions. Solid lines represent fits to the second-moment ($D_{4\sigma}$) beam diameters according to the ISO 11146 standard. Inset: spatial beam profile.}
\label{fig:efficiency_M2}
\end{figure}

The experiment is implemented using the same configuration. Figure~\ref{fig:efficiency_M2}(b) shows the measured output signal energy and quantum efficiency as a function of input pump pulse energy. Maximum conversion is reached at a pump energy of 1.9 \si{\micro\joule}, where the signal energy is measured to be 0.8 \si{\micro\joule}, yielding a pump-to-signal efficiency of 42\% and a quantum efficiency of 64\%, in good agreement with the simulated values of 44\% and 67\% respectively. The experimental data obtained at a pump energy of 1.9~\si{\micro\joule} are compared with simulation results at 1.65~\si{\micro\joule}, with both operating points corresponding to similar locations along the efficiency curve. This discrepancy could be attributed to non-ideal spatial and temporal beam profiles, which are not accounted for in the simulations. The optical losses accumulated through the multipass system arise from six reflections on the folding mirrors, residual absorption in the crystals, and the anti-reflection coatings of the LBO and $\alpha$-BBO crystals. They are measured to be 11.9\% for the pump and 9.6\% for the signal. To provide a more physically transparent figure of merit that isolates the parametric conversion process from these passive losses, the ratio of output signal photons to the number of pump photons that would be transmitted through the system in the absence of any conversion is 73\% at the optimal operating point, demonstrating that most pump photons surviving linear propagation losses are effectively converted into signal photons. The level of optical parametric generation (OPG), measured in the absence of seed, is also reported in Fig.~\ref{fig:efficiency_M2}(a), and remains low at the pump energy corresponding to optimal conversion, confirming that the amplified signal is seeder-driven and not contaminated by spontaneous parametric fluorescence. This conclusion is further supported by the measured pulse-to-pulse stability of the amplified signal, which amounts to 0.3\% rms. This value is identical to the pulse-to-pulse stability of the pump. Figure~\ref{fig:efficiency_M2}(b) shows the output signal energy and quantum efficiency as a function of CW seed power at the optimal input pump energy of 1.9 \si{\micro\joule}. The CW power that is necessary to reach full saturation of the amplification process is around 15 mW.

The spatial quality of the amplified beam is characterized at maximum extraction efficiency. Since the CW seed background is superimposed on the amplified pulsed signal on the camera, the unamplified CW contribution is systematically subtracted from all captured spatial profiles prior to analysis, isolating the amplified signal beam. The resulting beam quality measurement yields $M_{x\times y}^2 = 1.35\times 1.23$, as shown in Fig.~\ref{fig:efficiency_M2}(c), confirming the near-diffraction-limited performance predicted by the simulations. 

\begin{figure}[h]
\centering\includegraphics[width=0.95\linewidth]{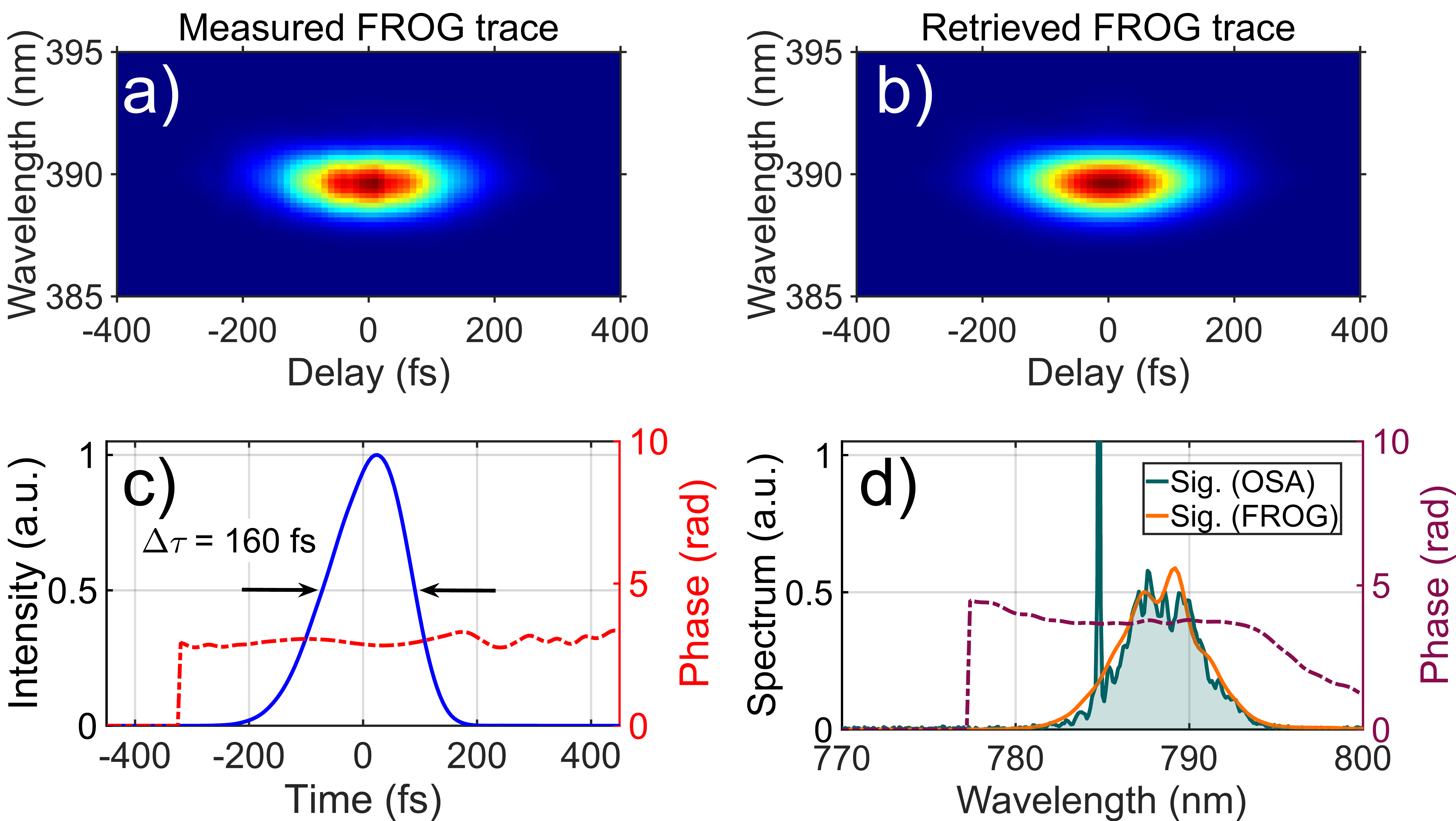}

\caption{Spectro-temporal characterization of the amplified signal at a pump pulse energy of 1.9 \si{\micro\joule}. \textbf{(a)} Measured FROG trace. \textbf{(b)} Retrieved FROG trace. \textbf{(c)} Retrieved temporal profile (blue solid line, left axis) and temporal phase (red dashed line, right axis), indicating a pulse duration of 160~fs FWHM. \textbf{(d)} Measured optical spectrum from the OSA (dark cyan solid line, left axis), compared with the spectrum retrieved from the FROG measurement (orange solid line, left axis). The corresponding retrieved spectral phase is shown (dark purple dashed line, right axis).}

\label{fig:spec_temp}
\end{figure}

The spectro-temporal quality of the amplified pulses is assessed through SHG-FROG measurements, shown in Fig.~\ref{fig:spec_temp}. The reconstructed temporal profile reveals a high-fidelity pulse with a FWHM duration of 160 fs, with an essentially flat spectral phase, consistent with a near-transform-limited pulse. The FROG-reconstructed spectrum is compared against an independent measurement acquired with a calibrated optical spectrum analyzer (Fig.~\ref{fig:spec_temp}(d)), showing excellent agreement; the latter also captures the CW seed peak. The amplified signal is centered at 783 nm with a FWHM bandwidth of 5.9 nm, yielding a time-bandwidth product of 0.46, very close to the Fourier-transform limit of 0.44 for a Gaussian pulse. The measured peak power of 4.9 MW yields a parametric peak power gain of $1.1\times10^{8}$ (approximately 80 dB). Taken together, the spatial, spectral, and temporal characterizations are in excellent quantitative agreement with the numerical predictions, validating both the physical model and the design principles underlying the concept.

In summary, we have experimentally demonstrated a CW-seeded, collinear multipass optical parametric amplifier in which successive gain stages are interleaved with free-space sections for mode matching. Temporal synchronization between the pump and signal is managed using a dedicated delay unit, while the idler is suppressed after each stage to limit back-conversion. Enabled by these mechanisms, the system achieves a quantum efficiency of 64\% and a 42\% pump-to-signal energy conversion efficiency. Operating at a 500-kHz repetition rate, the OPA delivers approximately 80~dB of parametric gain while preserving an excellent spatial beam profile and spectro-temporal quality.

This architecture can be adapted to a variety of different source parameters and target wavelengths. Depending on the nonlinear crystal and pump pulse duration and energy, the pump beam size is limited by the damage threshold. This size is tightly related to the free-space distance needed between consecutive gain stage, with longer distances being necessary as the pump size increases. The nonlinear crystal length is essentially determined by pump-signal GVM, which in turn determines the length of the GVM compensation birefringent plate, and, together with the pump intensity, the single stage gain. Following these guidelines, the concept can be readily adapted for a wide range of seed type, pump energies and repetition rates, and should also allow efficient use of low-gain nonlinear crystals across diverse spectral regions. The combination of its simplicity, performance, and versatility makes it uniquely suited to drive numerous applications such as time-resolved spectroscopy \cite{maiuri_ultrafast_2020} and multiphoton microscopy \cite{helmchen_deep_2005}.   
\\\\
\textbf{Funding.}
Agence Nationale de la Recherche (ANR-24-CE92-0010-02 MILLSTREAMS, ANR-23-CE30-0037 MILPATS).\\\\
\textbf{Disclosures.}
A patent has been filed on some of the aspects described in this work.\\\\
\textbf{Data Availability Statement.}
Data underlying the results presented in this paper are not publicly available at this time but may be obtained from the authors upon reasonable request.

\bibliography{biblio2}

\end{document}

%% file: preamble.tex
\usepackage{amsthm}
\usepackage{mathtools}
\usepackage{physics}
\usepackage{xcolor}
\usepackage{graphicx}
\usepackage[left=23mm,right=13mm,top=35mm,columnsep=15pt]{geometry} 
\usepackage{adjustbox}
\usepackage{placeins}
\usepackage[T1]{fontenc}
\usepackage{lipsum}
\usepackage{csquotes}